\documentclass[aps,prl,superscriptaddress,twocolumn,reprint,amsmath,amssymb
]{revtex4}

\usepackage{array}
\usepackage{float}
\usepackage{amsmath}
\usepackage{setspace}
\usepackage{pbox}
\usepackage{graphicx,amsmath,amssymb,appendix,amsfonts,latexsym,color,dcolumn,bm,mathtools}

\makeatletter

\newif\ifstructure
\structuretrue

\newcommand{\beq}{\begin{equation}}
\newcommand{\eeq}{\end{equation}}
\newcommand{\bea}{\begin{eqnarray}}
\newcommand{\eea}{\end{eqnarray}}

\newcommand{\cm}{cm$^{-1}$}

\makeatother

\begin{document}
\title{Understanding Radiative Transitions and Relaxation Pathways in Plexcitons}
\author{Daniel Finkelstein-Shapiro \footnote{daniel.finkelstein@iquimica.unam.mx}}
%\ead{daniel.finkelsteinshapiro@chemphys.lu.se}

\address{Division of Chemical Physics and Nanolund, Lund University, Box 124,
221 00 Lund, Sweden}
\author{Pierre-Adrien Mante}
\address{Division of Chemical Physics and Nanolund, Lund University, Box 124,
221 00 Lund, Sweden}
\author{Sema Sarisozen}
\address{Department of Chemistry, Izmir Institute of Technology, 35430 Izmir,
Turkey}
\author{Lukas Wittenbecher}
\address{Division of Chemical Physics and Nanolund, Lund University, Box 124,
221 00 Lund, Sweden}
\author{Iulia Minda}
\address{Division of Chemical Physics and Nanolund, Lund University, Box 124,
221 00 Lund, Sweden}
\author{Sinan Balci}
\address{Department of Photonics, Izmir Institute of Technology, 35430 Izmir,
Turkey}
\author{T\~{o}nu Pullerits}
\address{Division of Chemical Physics and Nanolund, Lund University, Box 124,
221 00 Lund, Sweden}
\author{Donatas Zigmantas \footnote{Lead contact: donatas.zigmantas@chemphys.lu.se}}
\address{Division of Chemical Physics and Nanolund, Lund University, Box 124,
221 00 Lund, Sweden}

\begin{abstract}
Molecular aggregates on plasmonic nanoparticles have emerged as attractive systems for the studies of polaritonic light-matter states, called plexcitons. Such systems are tunable, scalable, easy to synthesize and offer sub-wavelength confinement, all while giving access to the ultrastrong light-matter coupling regime, promising a plethora of applications. 
However, the complexity of these materials prevented the understanding of their excitation and relaxation phenomena. Here, we follow the relaxation pathways in plexcitons and conclude that while the metal destroys the optical coherence, the molecular aggregate coupled to surface processes significantly contributes to the energy dissipation.
We use two-dimensional electronic spectroscopy with theoretical modeling to assign the different relaxation processes to either molecules or metal nanoparticle. 
We show that the dynamics beyond a few femtoseconds has to be considered in the language of hot electron distributions instead of the accepted lower and upper polariton branches and establish the framework for further understanding.
\end{abstract}
\maketitle

\section{Introduction}

%{ \let\thefootnote\relax\footnotetext{Lead Contact: donatas.zigmantas@chemphys.lu.se} }

Cavity quantum electrodynamics (cQED) has been a very
successful testbed for the quantum mechanics of light-matter interaction
\cite{Walther2006,Mabuchi2002,Haroche1989,Raimond2001}, and exhibits
phenomena of both fundamental and practical interest 
 \cite{Kasprzak2006a,Martinez-Martinez2018,Yuen-Zhou2016,Zhong2016,Feist2015,Schachenmayer2015,Herrera2016,Galego2015,Hutchison2013,Hutchison2012,Thomas2016,Wang2014a}.
 Dressing matter with the electromagnetic modes of a cavity results
in hybrid light-matter states (polaritons) with properties that are
only recently beginning to be exploited for chemistry applications,
such as photocatalysis \cite{Mandal2019,Manuel2019}, remote energy transfer \cite{Coles2014,Saez2018,Xiang2020,
Du2018,Zhong2017,Krainova2020,Herrera2016} and polaritonic chemistry \cite{Du2019,Fregoni2020,Ribeiro2018,Feist2018}. As the light-matter coupling
strength increases, new and interesting effects arise tied to non-zero
ground state occupation of the cavity, so that finding systems that
can push the coupling into stronger regimes are desirable to explore
new photophysics and photochemistry \cite{Torma2015,Johannes2018,Ebbesen2016}.
Condensed-matter cQED realizations are particularly attractive as
they reach the ultrastrong coupling regime at room temperature with
Rabi splittings in excess of 30\% as compared to the emitter frequency
\cite{Walther2006,Chen2017,Wersaell2017,Roller2016,Antosiewicz2014,PhysRevLett.97.266808,PhysRevLett.93.036404,Balci2016,Santhosh2016,Hertzog2019,Hutchison2012}.
This regime can be reached both by molecules placed in a microcavity
\cite{Hutchison2012,Ebbesen2016}, or in systems where the field has
been confined by a plasmonic resonance to sub-wavelength dimensions,
as is the case for plexcitons (plasmon+exciton) \cite{Johannes2018,Balci2016}.
Confining the light mode in a plasmonic nanoparticle, as opposed to
a microcavity or a plasmonic array, provides a notable advantage:
the systems can be synthesized in large numbers using synthetic chemistry
methods, and exist as colloidal suspensions in solution. Thus, they
provide polaritonic states in a beaker. 

There is an increasing complexity of the dissipative environment in
polaritonic systems as we move from atoms in cavities, to molecules
in microcavities, and finally to molecules coupled to plasmonic nanoparticles.
The wealth of dissipative processes when replacing a cavity with a plasmon
-- for example Landau damping, electron-electron (e-e) scattering
and electron-phonon (e-ph) scattering -- modulate the dynamics in
non-trivial ways that cannot be ignored \cite{Voisin2001}. A stronger
coupling to a dissipative environment, however, needs not be necessarily
detrimental, and there are several examples where harnessing it increases
device efficiency, and therefore dissipation should be understood and explored
\cite{Lin2013,Scully2011}.

Ultrafast pump-probe studies have reported the relaxation
times of the excitation \cite{Fofang2011,Balci2014,Hranisavljevic2002,Vasa2013,Wiederrecht2008,Hoang2016}
as well as the Rabi oscillations in the case of molecules coupled
to gold nanoslit arrays \cite{Vasa2013}. Unfortunately, currently
there is no consensus on the plexciton lifetimes and dephasing processes,
likely stemming from system to system differences, and non-trivial
excitation wavelength dependence. For example, Hranisavljevic \textit{et
al.} found an exceptionally stabilized excited state with a 300 ps
decay constant \cite{Hranisavljevic2002}, while Balci \textit{et
al.} measured decay constants close to 10 ps, which are pump wavelength
dependent \cite{Balci2014}, on similar systems consisting of cyanine
dyes adsorbed on Ag plasmonic nanoparticles. More importantly, an
understanding of the dissipation in terms of the physical mechanisms
in the molecular aggregates or the plasmonic nanoparticles is lacking,
so that rational and systematic improvements of these materials are
hindered. The studies of microcavity polaritons in the infrared region
(where molecular vibrations are dressed by the cavity modes) are more
mature, and can provide useful guidelines as to what differences we
might expect between molecular and atomic systems \cite{Xiang2018}.
Notably, previous studies underscore the importance of dark states
\cite{Xiang2018,Zhang2019}, although extrapolating the same physics
to plasmonic-based hybrids in the visible regime should be done with
care.
Recent transient absorption studies at visible wavelengths find that two-particle polaritonic states play an important role \cite{DelPo2020}, while a two-dimensional electronic spectroscopy (2DES) investigation has measured a relaxation pathway from upper to lower branch without the involvment of the molecular dark states \cite{Mewes2020}.  

In this work, we explore in detail the ultrafast dynamics of plexcitons
and corresponding dissipation mechanisms. We start from a description
of the plexciton linear absorption by extending the response function
formalism to non-Hermitian Hamiltonians. This forms the basis to interpret 2DES experiments and to
propose a physical origin of each relaxation component observed in
the plexciton, either to processes inside the J-aggregates or inside
the metal nanoparticle.

\section{Results}

J-aggregates (TDBC, 5,5,6,6-tetrachloro-di(4-sulfobutyl) benzimidazolocarbocyanine),
Ag nanoprisms and plexcitons (TDBC-Ag nanoprism) were prepared as
reported previously (see Methods section, \cite{Balci2013,Balci2016}.).
The samples were prepared in solution and the reported measurements
were done in transmission mode.

\subsection{Linear response}

The absorption spectrum can be used to infer the homogeneous linewidth
(when it is much larger than the heterogeneous broadening) and the
energy structure of the first excited state manifold (Figure 1.a),
a prerequisite to interpret correctly the third order response signals
presented below. The absorption spectrum of the plexciton clearly shows the
upper and lower polaritonic branches (Figure 1.b).
The minimum of the dip in the spectrum is nearly coincident with the
J-aggregate absorption at 16,950 \cm (590 nm). \\

\begin{figure}[h]
\includegraphics[width=0.5\textwidth]{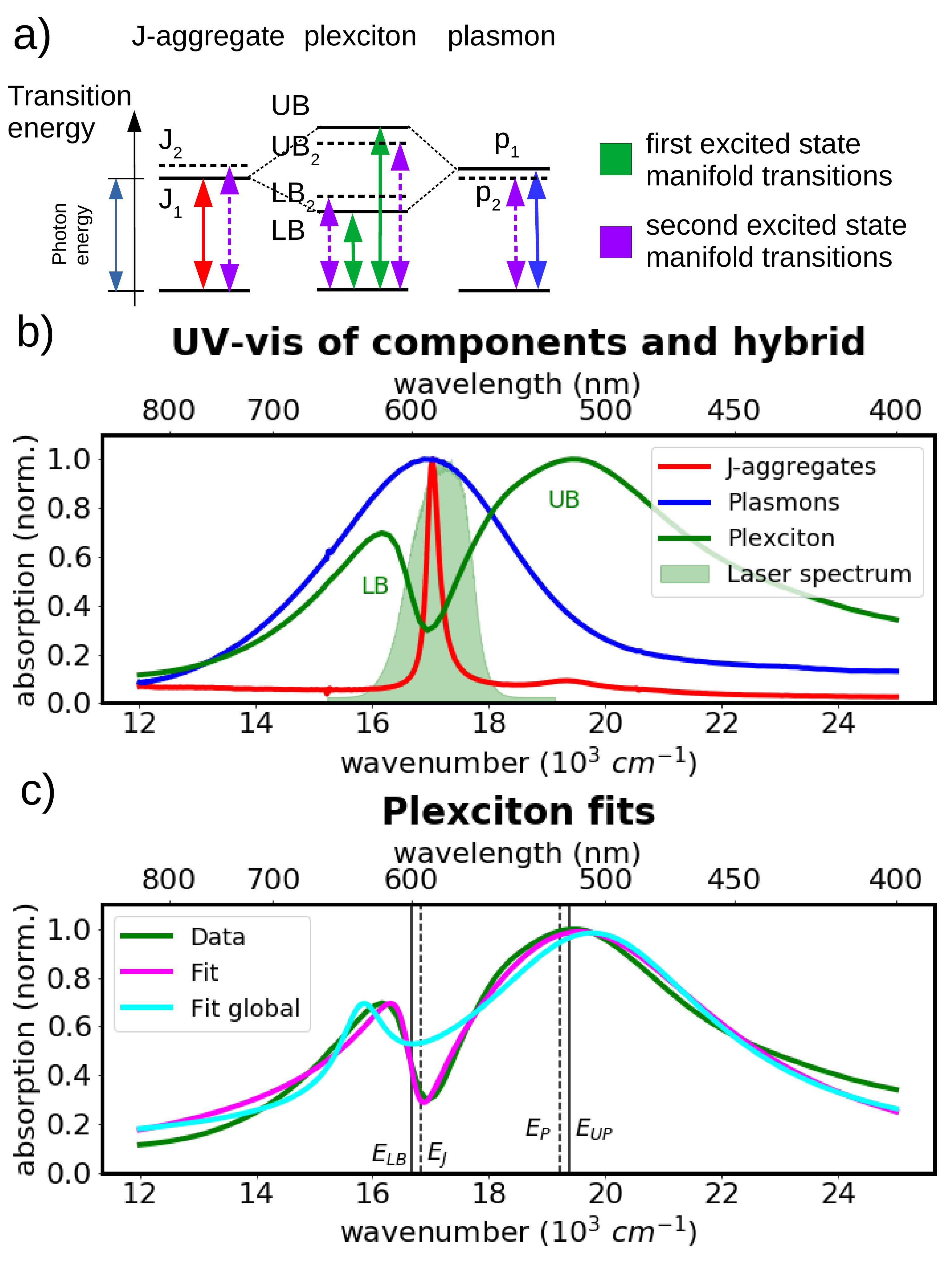}
\caption{\small Linear optical response. a) Transition energy of the ground to first and from the first to the second exciton manifolds. 
In plexcitons, matter and light states hybridize to give a lower and upper polaritonic branch. The shown nonlinearity between the first to second exciton manifold transition compared to the ground to first exciton manifold transition is due to a Rabi contraction. We neglect the transitions to higher electronic states lying two photon quanta above the ground state since their contribution is much smaller than double excitations of the levels lying one photon above the ground state. 
b) Absorption spectrum of J-aggregates
and extinction spectrum of plasmons and plexcitons, with the laser
spectrum of the pulses used in 2DES experiments. c) The measured and
fitted plexciton extinction spectra using the local and global approaches
(see main text). Dashed vertical lines indicate the uncoupled J-aggregate and
plasmon transition frequencies, while solid lines indicate the upper and lower polariton transition frequencies (we use $\omega=2\pi\tilde{\nu}c$).}
\label{fig:Figure1}
\end{figure}

In the following, through fitting of the lineshape we obtain the energy
and dephasing time of the first excited states manifold and motivate
the structure of the higher excited states manifold based on previous
measurements and theoretical models (Figure 1.a
and \cite{Agranovich1974,Minoshima1994}). An ideal J-aggregate consisting
of $N$ monomers with individual transition dipole moments $\mu_{0}$
features a single bright exciton state carrying all of the transition
dipole moment strength $\mu_{\text{bright}}=\sqrt{N}\mu_{0}$, and
$N-1$ dark states. Although this is strictly true only in the absence
of disorder, its 
 predictions are in agreement with the absorption
spectrum of the J-aggregate consisting of a very strong and narrow
absorption line. It is well-fit by a Lorentzian with center frequency
$\tilde{\nu}_{J_{1}}=17.04\times10^{3}$ \cm and full-width at half
maximum (FWHM) of $\gamma_{J_{1}}=120$ \cm, corresponding to a dephasing
time of 44 fs (Supplemental Figure S1). The double exciton states
manifold consists of a band whose transition from the first exciton
manifold ($1\to2$) is almost isoenergetic with the first transition
($0\to1$) \cite{Milota2013}, and its detuning can be calculated
from the linear Frenkel exciton
model \cite{Minoshima1994}. While
both first and second excited states form bands, we will refer to
them as single states under the viewpoint that most of the transition
dipole moment strength is carried by one state. % 
Analogously to the J-aggregate 
transition, we model the lineshape
of the plasmon  absorption in the Ag nanoparticles as a Lorentzian with
center wavenumber $\tilde{\nu}_{p_{1}}=16.92\times10^{3}$ \cm and
a width of $\gamma_{p_{1}}=1.78\times10^{3}$ \cm corresponding to
a dephasing time of 3 fs (Supplemental Figure S2). 

Given the electronic structure of the individual components, and the
expected coupling between them, we arrive at the energy level structure
shown in Figure 1.a. We summarize the state
energies and couplings in the following Hamiltonian, which is analogous
to that proposed by others \cite{Delga2014,Shah2013,Zelinskyy2012,Varguet2019, Piryatinski2020} (see Figure 1.a).
The Hamiltonian in the \{$g$, $p_{1}$, $J_{1}$
\} basis (that we call here site basis) reads: 
\begin{equation}
\begin{split}H_{\text{site}}=\begin{bmatrix}0 & F\mu_{gp_{1}} & F\mu_{gJ_{1}} \\
F\mu_{p_{1}g} & \hbar\bar{\omega}_{p_{1}} & V_{1} \\
F\mu_{J_{1}g} & V_{1}^{*} & \hbar\bar{\omega}_{J_{1}} \\
\end{bmatrix}\end{split}
\label{eq:NH-Hamiltonian}
\end{equation}
where $\bar{\omega}_{j}=\omega_{j}-i\gamma_{j}$ is a complex frequency
that accounts for the transition frequency $\omega_{j}$ and its associated
dephasing rate $\gamma_{j}$, $\mu_{ji}$ are the transition dipole
moments from $i$ to $j$, $F$ is the field strength and $V_{i}$
is the dipolar coupling between the J-aggregate and the plasmon's
$i$-th excited state. $g$ is the shared ground state, and $p_{i}$
and $J_{i}$ are the $i$-th excited state of the plasmon and J-aggregate,
respectively (the measured spectra are displayed in wavenumbers $\tilde{\nu}=\omega/(2\pi c)$
where $\omega$ is the angular frequency and $c$ is the speed of
light). The second excitations of the individual components are also labeled in Figure 1.a. 

In order to use the same theoretical approach for
the linear and third order response, we simulate the absorption spectrum using the response function formalism \cite{Mukamel1995} with two different limits of the Markovian bath \cite{Hofer2017}.
 In the polariton basis - obtained by diagonalizing $H_{\text{site}}$
at zero-field ($F=0$) and without including the dephasing (that is
setting $\bar{\omega}_{j}=\omega_{j}$), we can express the linear
absorption as a sum over polariton state contributions \cite{Cho2009,Maly2018}:
\begin{equation}
I(\omega)=\int dte^{i\omega t}\sum_{a=\text{UB,LB}}\mu_{ga}^{(g)}\mu_{ag}^{(g)}e^{(-i(\omega_{a}^{(g)}-\omega_{g})-\gamma_{a}^{(g)})t},\label{eq:global}
\end{equation}
where $\mu_{ag}^{(g)}$, $\omega_{a}^{(g)}$ and $\gamma_{a}^{(g)}$
are the transition dipole moment from level $g$ to the polaritonic
upper or lower branch $a\in\{LB,UB\}$, the transition frequency between
$g$ and $a$ and the dephasing rate, respectively. The $(g)$ or
$(l)$ (used below) superscript refer to the global and local approach taken to obtain the polariton branches. In the global
approach, the transition frequencies and transition dipole moments
are obtained from the diagonalization of the Hamiltonian without dephasing,
with the dephasing rate of each branch introduced phenomenologically
in the expression for the absorption (Eq. \eqref{eq:global}). This model
fits the spectrum poorly (cyan, Figure 1.c).
The global approach model is equivalent to fitting the absorption
spectrum to a sum of Lorentzians (here two) that in this case do
not correspond to the lineshape measured. We instead diagonalize the non-Hermitian
operator $H_{\text{site}}$ including the dephasing rates $\gamma_{J_{1}},\gamma_{p_{1}}$
(as the imaginary part of the transition frequencies) to obtain the
new (complex) polariton branch energies ${\omega}_{a}^{(l)}$ and
their respective transition dipole moments ${\mu}_{ag}^{(l)}$. We
modify the expression for the linear response accordingly to: 
\begin{equation}
I(\omega)=\int dte^{i\omega t}\sum_{a=UB,LB}{\mu}_{ga}^{(l)}{\mu}_{ag}^{(l)}e^{-i({\omega}_{a}^{(l)}-{\omega}_{g}^{*})t},\label{eq:local}
\end{equation}
\begin{table}
\centering %
\begin{tabular}{|l|c|c|}
\multicolumn{3}{c}{a. Separate components in solution}\tabularnewline
\hline 
 & $10^{3}$ cm$^{-1}$  & fs \tabularnewline
\hline 
\textbf{J-aggregate}  &  & \tabularnewline
$\tilde{\nu}_{J1}$  & 17.04  & \tabularnewline
$\gamma_{J1}^{-1}$  &  & 44 \tabularnewline
\hline 
\textbf{Plasmon}  &  & \tabularnewline
$\tilde{\nu}_{P1}$  & 16.92  & \tabularnewline
$\gamma_{P1}^{-1}$  &  & 3 \tabularnewline
\hline 
\multicolumn{1}{l}{} & \multicolumn{1}{c}{} & \multicolumn{1}{c}{}\tabularnewline
\end{tabular}\\
\centering %
\begin{tabular}{|>{\raggedright}p{3cm}|c|l|c|c|}
\multicolumn{5}{c}{b. Hybrid system in solution}\tabularnewline
\hline 
 & \multicolumn{2}{c|}{Local} & \multicolumn{2}{c|}{Global}\tabularnewline
\hline 
 & \pbox{20cm}{ $10^{3}$ cm$^{-1}$}  & \pbox{20cm}{ fs }  & \pbox{20cm}{ $10^{3}$ cm$^{-1}$}  & fs \tabularnewline
\hline 
\pbox{20cm}{\textbf{site} \textbf{basis}}  &  &  &  & \tabularnewline
$\tilde{\nu}_{J1}$  & 16.83  &  & 16.08  & \tabularnewline
$\gamma_{J1}^{-1}$  &  & 34  &  & \tabularnewline
$\tilde{\nu}_{P1}$  & 19.22  &  & 19.52  & \tabularnewline
$\gamma_{P1}^{-1}$  &  & 1.6  &  & \tabularnewline
$V_{1}$  & 0.98  &  & 0.98  & \tabularnewline
\hline 
\pbox{20cm}{\textbf{plexcitonic} \\
\textbf{basis}}  &  &  &  & \tabularnewline
$\tilde{\nu}_{LB}$  & 16.67  &  & 16.42  & \tabularnewline
$\gamma_{LB}^{-1}$  &  & 15  &  & 11 \tabularnewline
$\tilde{\nu}_{UB}$  & 19.38  &  & 19.59  & \tabularnewline
$\gamma_{UB}^{-1}$  &  & 1.7  &  & 2.1 \tabularnewline
\hline 
\end{tabular}\caption{Fit parameters for the linear optical response. a) Parameters of the radiative transitions of the individual components
in solution (J-aggregates and plasmons measured separately). The absorption
lineshape is fit with a Lorentzian (see SI for fits). b) Parameters
for the radiative transition of plexcitons. The spectrum is fit using
equation \eqref{eq:global} (global approach) or \eqref{eq:local}
(local approach). The upper and lower branch properties are obtained
by the diagonalization of the Hermitian Hamiltonian of Eq. \eqref{eq:NH-Hamiltonian}
(without dephasing) for the global approach, and of the non-Hermitian
Hamiltonian (with dephasing) for the local approach. Dephasing of
the branches for the global approach is added in the calculation of
the absorption spectrum while for the local approach it is included
in the Hamiltonian.}
\label{table:widths} 
\end{table}

so that the dephasing rates are included in the complex
energies of the polaritonic states.
As we see from Figure 1.c in the
local approach fit (magenta line) the agreement with the model is
excellent, with deviations attributable to the simplifying assumption
of representing the plasmon lineshape by a Lorentzian far away from
the resonance (Supplemental Figure S2). The name of the models (local vs. global)
refer to the approximations implemented to reach the underlying Markovian
master equation \cite{Hofer2017}, and a more detailed discussion
of their physical meaning can be found in Supplemental Note  S1.

The parameters of the fits are reported in Table \ref{table:widths}., assuming that the principal contribution to the resonance width is homogeneous broadening. The plasmon dephasing time is extremely fast (1.6 fs), in agreement
with previous measurements \cite{Wright1994,Weick2005}. This fast
dephasing rate is inherited to the plexcitonic branches (1.7 and 15
fs for upper and lower branch respectively). The difference in dephasing
rates can be attributed to the upper branch having more of a plasmonic
character than the lower branch. This is clear from detuning between
the plasmon (19.22$\times10^{3}$ \cm) and the J-aggregate transition
(16.83$\times10^{3}$ \cm). The frequencies of the radiative transition
of the individual components, that is of the J-aggregates in solution,
and that of the plasmonic nanoparticles in solution, differ from the
frequencies obtained from the fits for the J-aggregate and plasmon
transitions in the hybrid system. When bound to plasmonic
nanoparticles, the J-aggregates exhibit a redshift of 210 \cm, attributable to structural perturbations of J-aggregates when binding to a metal nanoparticle. 
The plasmon resonance in the hybrid system was substantially blue shifted with respect to the bare nanoparticles due to small structural differences arising from the surface stabilization properties afforded by the different molecular covering (citrate and polyvinylpyrrolidone for Ag nanoprisms and TDBC for plexcitons). We observe that the frequency of the plasmon in the plexciton system blue-shifts over the course of several weeks. It occurs on a timescale much longer than a single experiment and we expect this to have a small effect on the dynamics observed \cite{Voisin2001}. 
From the absorption fits we can see that the
plasmon carries most of the transition dipole moment amplitude ($\mu_{J_{1}g}=0.05\mu_{p_{1}g}$),
justifying the approximation of \cite{Delga2014} of considering that
only the plasmonic transition couples to the external radiation field.
The dephasing times of the plasmon and J-aggregate in the hybrid system are
larger than for the individual components in solution (1.6 vs. 3 fs
for plasmon, 34 vs 44 fs for J-aggregates) hinting at a contribution
from interfacial processes.
The Rabi splitting value extracted from the measured plexciton spectrum is $12$\% of the molecular transition frequency, placing the system in the ultrastrong coupling regime with respect to the transition frequencies, but not the dephasing rates. Thus we do not expect Rabi oscillations in the spectra altough the splitting is apparent. 

With this understanding of the energy states and their lineshapes
we turn to the excitation dynamics picture provided by 2DES. \\

\subsection{Time-resolved signals}

We measure the 2D spectra of the three systems at different population times
(referred here as $t_{2}$ \cite{Jonas2003}) to obtain the excited-state
dynamics of J-aggregates and plasmons, as well as plexcitons in order
to disentangle the contributions to dissipation from the metalic
and molecular components in the hybrid system (Figure 2).
\\

\textbf{J-aggregates}. The 2D spectra of J-aggregates
shows a broad excited-state absorption contribution (ESA, negative,
blue) almost coincident with a ground-state bleach (GSB, positive,
red) and stimulated emission signal (SE, positive, red), indicating
a near isoenergetic gap between the ground state and first excited
state band, and between the first and second excited state bands (see
Figure 2, Supplemental Note S2 and Supplemental Figure S3). This is in
accordance to the previous measurements \cite{Lee2001,Sundstrom1988,Minoshima1994}
and justifies the energy level scheme of Figure 1.a.
The population of the excited state decays with very little spectral
shift and can be fitted with three lifetimes of $80\pm20$ fs (43\%), $4.2\pm0.4$
ps (47\%) and a very long time, corresponding to excited state lifetime ($>100$ ps, 10\%). 
Those agree with the time constants obtained from previous transient absorption
experiments, that have assigned the first component to relaxation
into dark exciton states, the second component -- to exciton-exciton annihilation
(few ps, which depends on the excitation density) and the long-lived
component -- to relaxation back to the ground state \cite{Lee2001,Sundstrom1988,Minoshima1994}. The amplitude of the fastest component (close to half of the total signal) is consistent with the loss of the stimulated emission pathway during the waiting time $t_2$ due to the transition to dark states. In J-aggregates the dark states lie above the bright   
 transition so that only the levels within a thermal energy are accessible. \newline 

\textbf{Plasmons.} The 2D spectra of Ag nanoparticles show a single nodal
line delimiting negative and positive spectral regions (Figure 2).
The dynamics are characterized by spectral shifts and sign reversals
across this nodal line, and can be separated into the early time dynamics
($t_{2}<4$ ps), dominated by electron-electron (e-e) and electron-phonon
(e-ph) scattering and late time dynamics ($t_{2}>4$ ps) corresponding
to hot lattice ground state relaxation 
\cite{Hu2003,Hartland2011,Brown2017,Voisin2001,Lietard2018}. We can
think of the early time dynamics as corresponding to the processes
that transfer the energy from the initial
electronic excitation  to 

\begin{widetext}

\begin{figure}[h!]
\includegraphics[width=0.95\textwidth]{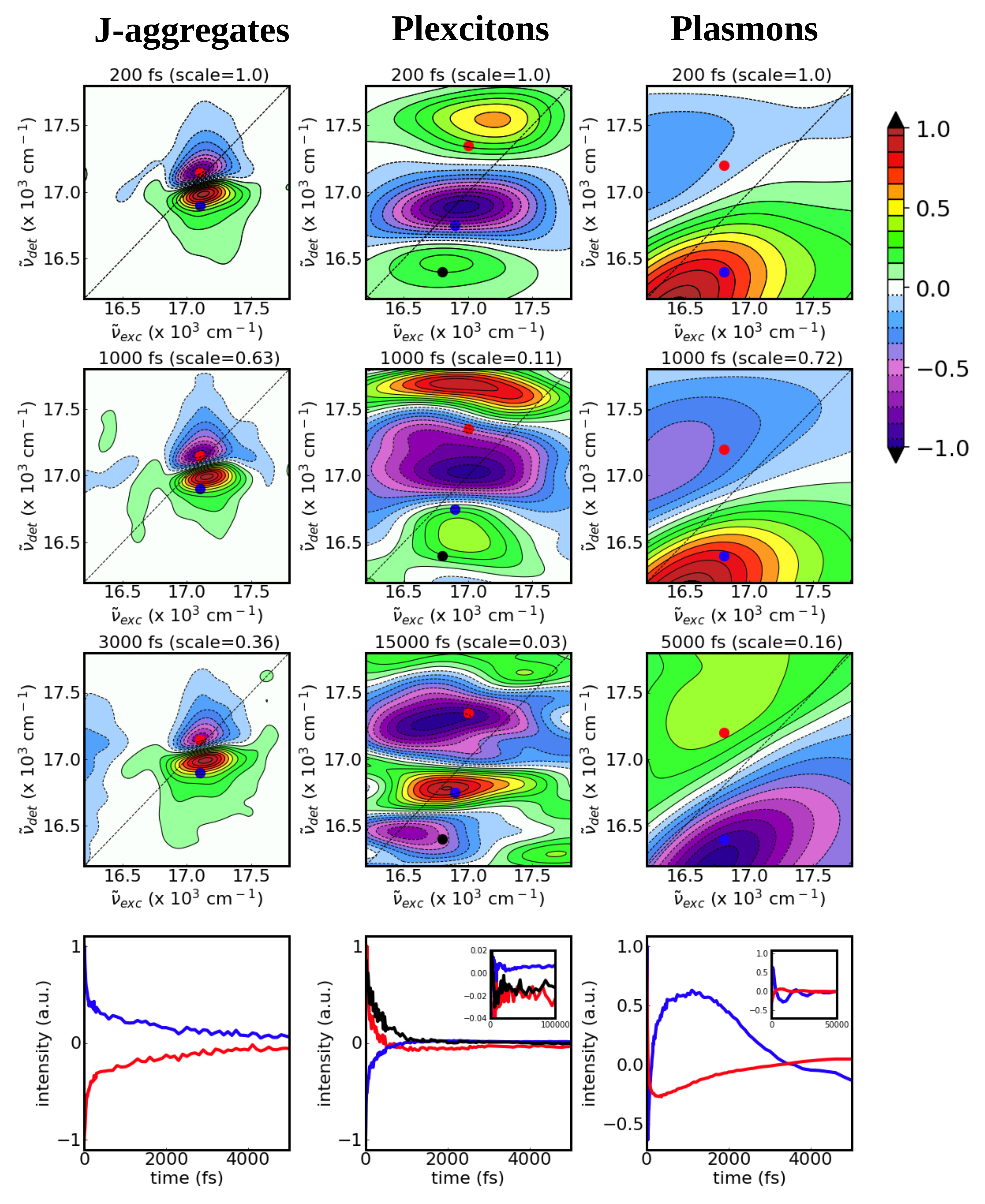}
\caption{\small 2D spectra of J-aggregates, plasmons and plexcitons. 2D spectra at different population times $t_{2}$, as well as amplitude time
traces for J-aggregates, plexcitons and plasmons (lowest row) at the points indicated in the 2D spectra above. Insets
for plexcitons and plasmons show the late-time signal. The data is
shown in normalized color scale with the scaling factor shown in parenthesis.}
\label{fig:Figure2}
\end{figure}

\end{widetext}

\noindent the lattice, and the late time dynamics -- as the response of the lattice
to the excess energy, notably lattice expansion due to the increase in temperature and the excitation of acoustic
phonon modes \cite{Voisin2001}. We start the analysis of the plasmon data at $T=100$ fs after the spatial modulation of the index of refraction imposed by the exciting pulses  has disappeared \cite{Lietard2018}.
The evolution of the signal can be very well fit by a phenomenological equation analogous
to the two-temperature model (see \cite{Sun1994,Link1999} and Supplemental Note S3, Supplemental Figures S5-S9): 
\begin{equation}
\begin{split}
I(t) & =A_{1}(1-e^{-t/\tau_{ee}})e^{-t/\tau_{eph}} \\
&+A_{2}(1-e^{-t/\tau_{eph}})e^{-t/\tau_{phph}} \\
&+A_{3}(1-e^{-t/\tau_{eph}})e^{-t/\tau_{vib}}\cos(\omega_{0}t-\phi) \\
&+A_{4}(1-e^{-t/\tau_{eph}})e^{-t/\tau_{vib}}
\end{split}
\end{equation}
where $\tau_{ee}=300\pm200$ fs, $\tau_{eph}=1.4\pm0.2$ ps are the
inferred e-e and e-ph scattering times, $\omega_{0}=2\pi/17.0$ rad/ps$^{-1}$
is the angular frequency of the phonon (breathing mode), $\phi=25^{o}$ --
its phase, and $\tau_{vib}=15$ ps -- its dephasing time. The thermalization
time of the silver nanoparticle with the environment on the timescale of
the phonon-phonon scattering time $\tau_{phph}$ is longer than the
measured population time range ($>$ 1 ns). 
The asymptotic phase $\phi$ of the breathing mode carries information
on the excitation mechanism of the acoustic phonon mode. The extracted
$25^{o}$ phase is in a reasonable agreement with the thermal expansion
mechanism (predicted at 
$\phi=\arctan(\omega_{0}\tau_{e-ph})=28^{o}$) and constitutes a signature
that energy is dissipated inside the metal nanoparticle, as opposed to evidence of a strong excited electron pressure (see Supplemental Note S3.3) \cite{Voisin2000}. \\

\textbf{Plexcitons}. 
The early time 2D spectra of the plexcitons consists of two positive
regions identifiable with the upper and lower branches, flanking a negative feature (Figure 2).  The time evolution of the signal can be well fit by the sum of four exponential decays with time constants of $\tau_d = 40\pm20$ fs, $\tau_{ee} = 400\pm100$ fs and $\tau_{eph} = 1.6\pm0.6$ ps and $\tau_{\text{solvent}} > $1 ns, where the last signal appears as a constant background at long times in our measurements. 
The shortest time constant carries the largest amplitude ($>$ 50 \%) and corresponds to a pure decay without any appreciable spectral diffusion (see Supplemental Note S4 and Supplemental Figures S10-13). The second and third time constants correspond to pronounced spectral shifts and the 2D spectra show little evolution after this. The time traces (Figure 2, last row) also exhibit fast oscillations on top of the exponential decays,
with dominant wavenumbers of 675 \cm and 1200 \cm . These correspond
to the Raman modes observed in the Surface-Enhanced Raman Spectroscopy
measurements of the same plexciton system (see Supplemental Figure
S14, \cite{Balci2013}). We leave a detailed discussion of the mechanisms behind each time constant for after an analysis of the simulations of the 2DES experiments. \\

\textbf{Simulations of the plexciton 2D spectra}. We simulate the
third-order response for very early times and very late times (Figure 3,  Supplemental Note S5 and Supplemental Figures S15-20 for details
of the simulations). For very early times (i.e. at times much shorter than the e-e scattering time), we consider that the optical response contribution from the non-equilibrium electrons inside the metal is negligible. At very late times when all excitations have decayed back to the electronic ground state, all the excitations of the system happen from the equilibrium Fermi distribution. The intermediate
times are not simulated here, and require an involved model based
on Boltzmann equations, which is beyond the scope of the current study.

At early times, we follow work on infrared cavities \cite{Ribeiro2018} and model the nonlinearity of the plexcitons as a Rabi contraction of the excited state, where the probe pulse exciting a system with a fraction $f$ of the total number of molecules per 
nanoparticle $N$ already excited and a smaller ground state Rabi splitting $\propto \sqrt{(1-f)N}$. We find that we need to account as well for excitation induced dephasing (EID) and excitation induced shifts (EIS). We neglect the contributions of Feynman diagrams that evolve in the excited state 
during $t_2$ as the polariton branches decay into intraband metal excitations or molecular dark states shortly after excitation, and these do not have radiative transitions. The late-time simulations assume that the probe pulse examines an electronic 
ground state with a hot lattice and a consequent redshifted plasmon transition, as has been reported for bare metals \cite{Voisin2001}. We find that while this model reproduces the positive and negative nodes of the spectrum, lower values of the plasmon transition need to be used (Details of the simulations are found in the Supplemental Note S5.)

We first notice that the absence of the upper and lower branch structure along the excitation dimension, both at early and late times, is well explained by the finite bandwidth of the pulses. 
There is a large parameter space that can explain the experiments both in the local and global approaches, so that experiments with broadband pulses will be needed in the future studies in order to extract unambiguous values of the parameters. 
Nonetheless, simulations support our assignment of the upper and lower polariton branches as the two positive signals at early times. 
\begin{widetext}

\begin{figure}
\includegraphics[width=1.0\textwidth]{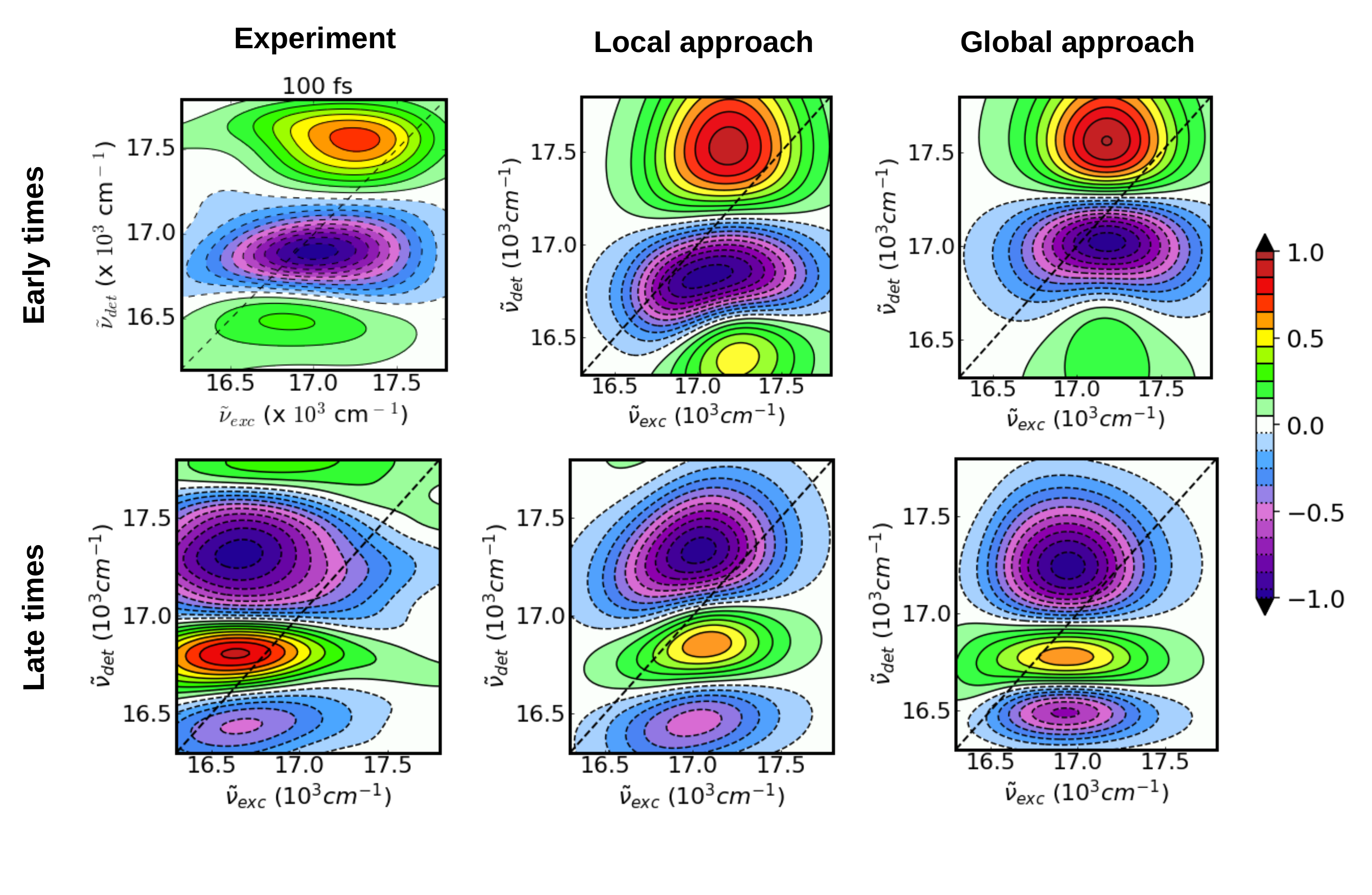}
\caption{2DES simulations. Measured and simulated 2D spectra of plexcitons at early ($t_{2}=0$) and late times. The simulations were carried out both in the local and global approach. Although both approaches give a similar fit to the 2D spectra, the local approach fits the linear absorption much better.}
\label{fig:Figure_3}
\end{figure}

\end{widetext}
Similar to modeling of linear absorption, sharper features can be seen with the local approach along the excitation dimension, as is apparent in the broadband simulations (see Supplemental Figures S17 and S18). 
At late times, our simulations suggest the presence of a hot ground state. In this case, only the local approach reproduces the positive and negative features observed in the experiment (2 positive, 2 negative) while the global approach can only explain some (2 negative, 1 positive, see Supplemental Figures S19 and S20).

\section{Discussion}

We now analyze each of the four components of the plexciton decay
dynamics (labeled d, ee, eph, and ph-ph)
and justify their assignment (see Figure
4). Our analysis is based on the relaxation
time constants, the 2D spectral signatures and the overall cohesiveness
of the proposed dissipation mechanisms. We notice that the dephasing rate of the optical coherence is not measured
directly, but inferred from the
absorption lineshape, as discussed in the first section of the manuscript.

% Explanation of the different time constants
The signature of the first time constant $\tau_d = 40$ fs is that of pure decay without any spectral shift and appears mainly in the upper polariton branch region (see Supplemental Figure S11). The J-aggregate has an ultrafast component $<100$ fs that does not cause any spectral shift either (corresponding to transfer to dark states) and both simulations as well as experiments in microcavities have noted that this very fast transfer to dark states occurs mostly from the upper branch, as we also observe \cite{Agranovich2003,Virgili2011,Groenhof2019}. However, the amplitude of this fast component in plexcitons exceeds 50\% of the total signal so that it cannot be explained exclusively by transfer to dark states which would entail a suppression of the SE pathway, but likely includes some relaxation of electrons to the ground state. 
In plasmons, a spatial long range order imposed by the laser sequence produces a strong distortion and variation of the signal at very early times \cite{Lietard2018}. In our plasmonic nanoparticles the first 100 fs are dominated by this spatial order, however the spectral distortion of

\begin{widetext}

\begin{figure}
\includegraphics[width=1.0\textwidth]{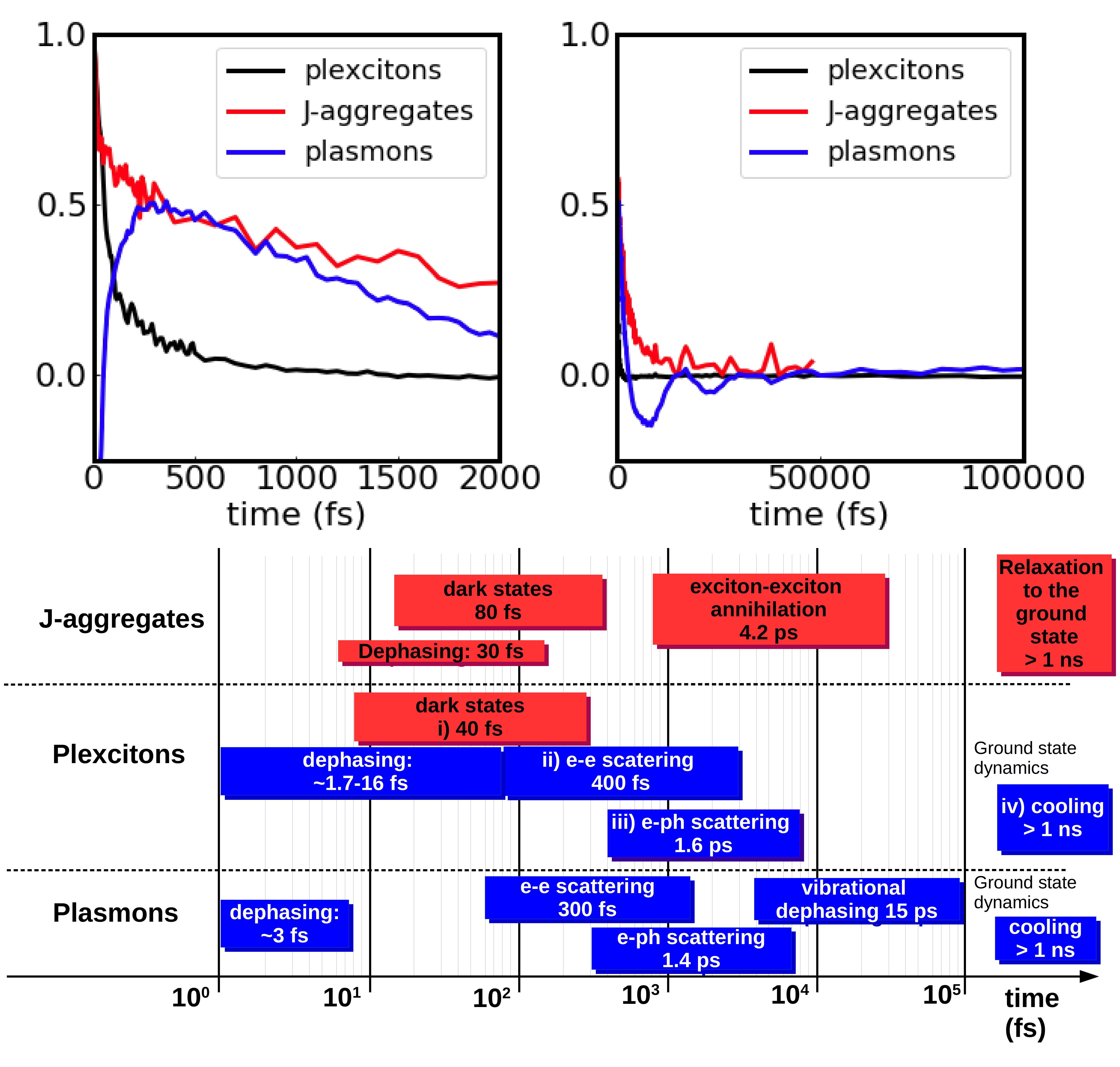}
\caption{Summary of dynamics. Top panel: representative kinetic traces of J-aggregates (marked by the blue dot of Figure 2, J-aggregate spectra), plasmons (marked by the blue dot of Figure 2, plasmon spectra) and plexcitons (lower branch, marked by the black dot of Figure 2, plexciton spectra) at short and long times.  Bottom panel: summary of the time constants measured in the experiments. The bars for a given time constant $\tau$ are drawn from $\tau$/5 to 5$\tau$. Vibrational dephasing also exists in plexcitons, however the energy deposited inside the metallic nanoparticle is insufficient to launch an acoustic mode, so that the rate of this process cannot be measured in our plexciton experiments.}
\label{fig:Figure_4}
\end{figure}

\end{widetext}

this process are absent in plexcitons so that it is doubtful whether the same process occurs in the decorated nanoparticles. A third possibility is that of radiation damping of the plasmon transition that leads to recombination to the ground state. This mechanism is likely to be unchanged in bare or TDBC-decorated Ag nanoparticles and cannot account for the loss of the acoustic mode vibration (explained below). 
We thus propose that this fast decay component corresponds to a combination of radiation damping and transfer to J-aggregate dark states, predominantly from the upper branch.  
During the second component $\tau_{ee}$ (400 fs) we see a pronounced spectral shift. The fact that J-aggregates have no processes that induce a spectral shift, while plasmons do, and with a similar time constant (300 fs), suggests that this process corresponds to the e-e scattering inside the metal nanoparticle. Unlike EIS, mentioned previously and which occurs almost instantaneously, the shift discussed here develops on the timescale of electron electron scattering.
The constant background at very long times is assigned to a hot lattice plexciton signal (see Figure 3, Supplemental Note S5 and Suppemental Figures S13, S19 and S20), so that the process that extends beyond the time range of the instrument ($>1$ ns) is likely thermalization of the solvent via phonon-phonon scattering with the time $\tau_{ph-ph}$. Previous measurements set this process in the hundreds of ps \cite{DelFatti2000}.
The third component $\tau_{eph}$ (1.6 ps) induces a small spectral shift. The time constant for this process is close to the time constant for e-ph scattering measured in the Ag nanoprisms (1.4 ps) and suggests the same underlying process. This tentative assignment is bolstered by the presence of a hot lattice after this process is over. Alternate assignments of this third time constant would  require an alternate mechanism for warming the lattice, however warming via the adsorbed molecules is not likely as this process is governed by $\tau_{ph-ph}$ and would also require several hundreds of ps. 
It is also striking that the replenishing of the lower polariton by the dark state reservoir is absent (contrary to microcavities \cite{Virgili2011}) and that the long lived dark modes do not survive after a few ps. In fact, the lower polariton dynamics seem to be determined by $\tau_{ee}$ and $\tau_{eph}$, and we attribute this to the direct coupling between molecules and the metallic surface which likely connects the relaxation of TDBC dark states with that of the non-equilibrium electron distribution inside the lattice. 
Lastly, we mention that a signature of exciton-exciton annihilation -- corresponding to a pure decay of the signal -- is not observed, although we cannot rule out that it is a process that contributes to the third time constant.

Our measurements suggest that enough energy is dissipated outside
of the metal nanoparticle so that the acoustic phonon cannot be excited
to a detectable intensity. It is most likely dissipated in the J-aggregate with possible contributions from the surface, given the fast transfer rate to dark states, which can compete with e-e and e-ph scattering. While the coherence of the plexciton is limited by processes inside the metal, the energy dissipation would thus be strongly influenced by ultrafast processes inside the J-aggregate. This indicates that to increase the lifetime of the excitation we should first modify the molecular component, for example by going towards the single molecule limit where the number of dark states is significantly reduced, and isolating the molecules from the metal so that recombination via surface processes is suppressed.

Furthermore, the observation that e-e scattering events exist in
plexcitons reveals a crucial point in plasmon-based polariton systems:
that one can no longer speak exclusively in terms of upper or lower
branches once dephasing processes have turned the initial coherent
excitation into a non-equilibrium distribution of electrons and holes.
A framework beyond Rabi type Hamiltonians, including the continuum
states of the metallic band is needed in order to describe anything
other than the early or late times.
Our findings are different from microcavity dynamics as much of the signal arises from the non-equilibrium distribution of electronic excitations inside the metal. 
Whereas the direct interaction between molecules and silver surface excitations of microcavities is virtually inexistent (because the mirrors are usually coated by an insulating polymer layer and the cavity mode antinode lies inside the cavity), there is strong coupling between molecules and the silver surface in our samples that is beyond the dipolar coupling of the bright molecular and plasmon transitions as evidenced by the short lifetimes of the TDBC dark states.
Thus metallic excitations can directly exchange energy and charge with adsorbed molecules. The possibilities afforded by this interaction in terms of hot electron chemistry set apart plasmon- from microcavity-based polariton states. 

Many possible applications of atomic cQED systems have been explored
throughout the years, from basic tests of quantum mechanics and quantum
information processing to studies of strong light-matter coupling,
its usefulness has been proven many times over. In contrast, the field
of molecular cQED systems -- and more so when the cavity is substituted
by a plasmonic mode -- is relatively recent. We are still learning
what these systems are capable of and in what ways they differ from
the more traditional cQED versions. There are many interesting phenomena:
delocalization, collective behavior, chemical reactivity and remote
energy transfer \cite{Kasprzak2006a,Martinez-Martinez2018,Yuen-Zhou2016,Zhong2016,Feist2015,Schachenmayer2015,
Herrera2016,Galego2015,Hutchison2013,
Hutchison2012,Thomas2016,Wang2014a, 
Coles2014,Saez2018,Xiang2020,
Du2018,Zhong2017,Krainova2020,Du2019,
Fregoni2020,Ribeiro2018,Feist2018}. 
In our work, we show that we need to understand the basic excitation
dynamics occurring in metals when these are strongly coupled to molecules.
The stronger coupling of plasmons to their baths
than to the J-aggregates means that the site states are dressed by
the dissipative modes. The resulting spectrum and underlying phenomenon
is more akin to an interference-type process than a Rabi splitting
\cite{FinkelsteinShapiro2020}. This type of physics has been recognized
before in scattering settings \cite{Faucheaux2014,Pelton2019}, however,
it has not been connected to the local vs. global approach to Markovian
master equations. As we show in this work, taking the local approach
requires an extension to non-Hermitian Hamiltonians in the response
function formalism. Thus our study provides both the motivation and
groundwork to study the nonlinear response of Markovian systems in
the limit of large dissipation that dominates plasmonic materials.
We believe that given the short coherence times in colloidal suspensions
of localized surface plasmon materials, a very interesting application
of plexcitons arises in the manipulation of optical transitions to
favor charge-transfer to an acceptor, or to promote catalysis, as
opposed to performing coherent operations with these states.

\section{Conclusion}

We have presented a systematic study of plexciton
dynamics using linear and third-order optical response. We are able to explain
the lineshapes using a non-Hermitian Hamiltonian extension of response
function formalism, which are rooted in the local approach to Markovian
master equations. We have tracked the excitation evolution and identified
one dephasing and four relaxation timescales which we have attributed
to either molecular-like or metallic-like processes. Remarkably, we
find that a significant part of the energy is dissipated outside of
the metal lattice. Our work suggests that when considering the relaxation
of plexcitons a more nuanced version of the dynamics in terms of hot
electron distributions is needed.

Realizations of polaritonic systems continue to bring new insights
into basic quantum physics concepts and emergent behavior of hybrid
systems. Because the plasmon-based polaritons exists at the intersection
of strong-light matter coupling and important condensed matter and
chemical processes, such as interfacial charge transfer and photocatalysis,
they are uniquely fit to bring desirable features of the former into
the societally relevant applications of the latter.

\section{Methods}

\textbf{Synthesis.} We synthesized plasmonic and plexcitonic nanoparticles
as reported in our earlier studies \cite{Balci2013,Balci2016}. Briefly,
the Ag nanoprisms with varying edge lengths were wet-chemically synthesized
by using a seed-mediated protocol. Initially, isotropic seed Ag nanoparticles
were synthesized, which was then followed by a slower growth of anisotropic
Ag nanoprisms. By controlling the number of seed nanoparticles in
the growth solution, the size of the nanoprism can be easily tuned.
Concurrently, the localized surface plasmon polariton frequency of
the nanoparticles vary from 400 nm to 1100 nm. Plexcitonic nanoparticles
used in this study were synthesized by self-assembly of a J-aggregate
dye (TDBC, 5,5,6,6-tetrachloro-di(4-sulfobutyl) benzimidazolocarbocyanine,
FEW Chemicals) on Ag nanoprism surfaces. It should be noted that the
excess dye molecules were removed by centrifugation. \\

\textbf{Two-dimensional electronic spectroscopy.} The 2DES spectrometer
is described in detail in \cite{Augulis2011}. Briefly, femtosecond
broadband pulses were generated by feeding the 1,030 nm output of
a Pharos laser (Light Conversion Ltd) to a commercial NOPA (Light
Conversion Ltd) to generate a pulse at 590 nm. The narrow-band spectrum
(see Supplemental Figure S15, simulations) where generated by compressing
the pulse using chirped mirrors and a prism compressor, and subsequently
limiting the spectrum using a blade edge to obtain a 40 nm FWHM pulse
of less than 25 fs duration (used for J-aggregates and plexcitons
measurements). To obtain the broadband pulse we used a home-built
pulse shaper arranged in a folded 4f-geometry. The setup is based
on a dual liquid crystal spatial light modulator (SLM-S640d, Jenoptik)
enabling simultaneous shaping of phase and amplitude \cite{Wittenbecher2019}.
Second order dispersion from material in the beam path is for the
most part compensated for by a pair of chirped mirrors in combination
with a prism compressor consisting of two fused silica prisms at a
separation of 310\,mm. Higher order phase distortions are compensated
for by the pulse shaper. With this arrangement broadband pulses with
a duration of 11 fs were obtained (used for the Ag nanoparticles measurement).
All four beams are focused at the sample to the spot size of 160 $\mu m$.
The pulses were attenuated to pulse energies in the range of 0.1 to
1.25 nJ/pulse. Polarizations of pulses 1 and 2 were set to magic angle
with respect to pulse 3 and the local oscillator. Several measurement
series were carried out across population times. The coherence time
was scanned from -71.6 to 151.2 fs in 1.4 fs steps for J-aggregates,
from -42 to 63 fs for plasmons and from -50.4 to 50.4 fs for plexcitons.
The resolution of the measurement was 60 \cm at the detection frequency
and 290 \cm at the excitation frequency. \\

\section{Acknowledgments}

Funding is acknowledged from the European Union through the Marie
Sklodowska-Curie Grant Agreement No.702694, from Nanolund and from
V.R. grants 2017-05150, 2017-04344 and 2018-05090. We also acknowledge
stimulating discussions with Patrick Potts.

\section{Author contributions}

D.F.S. and D.Z. conceived the idea. D.F.S., L.W., I.M. and D.Z. designed
and performed the experiments. D.F.S., P.A.M. and T.P. discussed the
theory and D.F.S. carried out the simulations. S.S. and S.B. synthesized
the samples. D.F.S. and P.A.M analyzed the data and D.F.S. wrote the
manuscript, edited by P.A.M., T.P. and D.Z., with input from all the
other authors.

\section{Declaration of Interests}

The authors declare no competing interests.

\bibliography{plexcitons,Fano,ebbesen,J-aggregates,from_second_review}

\end{document}